\documentstyle[aps]{revtex}
\begin{document}

\author{Gabriele Migliorini$^{1,2}$ and A. Nihat Berker$^{1,2,3}$}
\address{
$^1$Department of Physics, Massachusetts Institute of Technology\\
                                Cambridge, Massachusetts 02139, U.S.A.\\
$^2$Feza G\"ursey Research Center for Basic Sciences\\
                                \c Cengelk\"oy, Istanbul 81220, TURKEY\\
$^3$Department of Physics, Istanbul Technical University\\
                                Maslak, Istanbul 80626, TURKEY}
\title{Finite-Temperature Phase Diagram of the Hubbard Model}
\maketitle

\begin{abstract}
The finite-temperature phase diagram of the Hubbard model in
$d=3$ is obtained from renormalization-group analysis. It exhibits, around half
filling, an antiferromagnetic phase and, between 30\%--40\% electron or hole
doping from half filling, a new $\tau $ phase in which the electron hopping
strength $t$ asymptotically becomes infinite under repeated rescalings. Next to
the $\tau $ phase, a first-order phase boundary with very narrow phase
separation (less than 2\% jump in electron density) occurs. At temperatures
above the $\tau $ phase, an incommensurate spin modulation phase is indicated.
In $d=2$, we find that the Hubbard model has no phase transition at finite
temperature.
\end{abstract}

\narrowtext

The Hubbard model\cite{Hubbard} is the bare-essentials realistic model of
electronic conduction, yet essentially no knowledge has existed even
phenomenologically on its most frontal macroscopic feature, namely its phase
diagram at finite temperatures. In this research, we obtain a finite-temperature
phase diagram for the Hubbard model in spatial dimension $d=3$, from an
approximate renormalization-group calculation with flows in a 10-dimensional
Hamiltonian space. This rich phase diagram, in the variables of temperature,
electron density, and on-site repulsion, exhibits, around half-filling, an
antiferromagnetic phase completely due to electron hopping. At 30--40\% electron
or hole doping from half filling, a new $\tau$ phase occurs with distinctive
conduction property. In the neighborhood of the $\tau$ phase, a phase
separation so narrow that the jump in electron density is less than 2\% occurs.
At temperatures above the $\tau $ phase, an incommensurate frozen spin
modulation phase is indicated. In $d=2$, no phase separation or other phase
transition occurs at finite temperature in the Hubbard model, in
contrast\cite{KIVEL,FB} to the closely related, but less realistic, tJ
model of electronic conduction.

The Hubbard model is defined by the Hamiltonian
\begin{eqnarray}
 -\beta{\cal H} & = & -t \sum_{<ij>,\sigma}
\left( c_{i\sigma}^{\dagger}c_{j\sigma}+ c_{j\sigma}^{\dagger}
c_{i\sigma } \right) \nonumber \\
&               &       -U
\sum_i n_{i\uparrow }n_{i\downarrow }+\mu  \sum_i n_i \,\,\, ,
\end{eqnarray}
where $c_{i\sigma }^{\dagger }$ and $c_{i\sigma }$ are
the electron creation and annihilation operators with spin $\sigma = \uparrow $
or $\downarrow$ at site $i$ of a cubic lattice, $<ij>$ indicates summation over
all nearest-neighbor pairs of sites, and

\begin{equation}
n_{i\sigma }=c_{i\sigma }^{\dagger}c_{i\sigma} \,\,
{\rm and} \,\,
n_i=n_{i\uparrow }+n_{i\downarrow} \,\,\,
\end{equation}
are
the electron number operators. The terms in the Hamiltonian of Eq.(1) are,
respectively, the kinetic energy term, the on-site repulsion $(U>0)$ term, and
the chemical potential term included in order to study the system over its
entire density range from zero to two electrons per site.

The renormalization-group transformation is formulated\cite{FB,FBT} by
first considering a $d=1$ system. An exact renormalization group transformation
can be formally written,
\FL
\begin{eqnarray}
&& <u_1u_3u_5... \mid e^{-\beta ^{\prime }{\cal H}^{\prime }}\mid
v_1v_3v_5...> \nonumber \\
&& \sum_{w_2w_4w_6...}
                <u_1w_2u_3w_4u_5w_6...\mid e^{-\beta {\cal H}}\mid
v_1w_2v_3w_4v_5w_6...> \,\,\, ,
\end{eqnarray}
where $u_1,w_2,v_3,$ etc.\ represent the single-site states.
Primes indicate the renormalized system. The transformation given in Eq.(3)
conserves the partition function, $Z=Z^{\prime }$, but cannot be implemented due
to the non-commutativity of the operators in the Hamiltonian. An
approximation is
used:
\widetext
\begin{eqnarray}
{\bf Tr}_{\rm even \,\, sites}
\exp (-\beta {\cal H})
 =  && {\bf Tr}_{\rm even \,\, sites}
                        \exp \left( \sum^{\rm even}_i  -\beta {\cal
H}(i-1,i)-\beta {\cal H}(i,i+1) \right)
\simeq  \prod ^{\rm even}_i
                        {\bf Tr}_{w_i}
                        \exp \left( -\beta {\cal H}(i-1,i)-\beta {\cal
H}(i,i+1) \right)  \nonumber \\
&& =  \prod^{\rm even}_i
                        \exp \left( -\beta ^{\prime }{\cal H}^{\prime
}(i-1,i+1) \right)
\simeq  \exp \left(
                        \sum^{\rm even}_i
                        -\beta ^{\prime }{\cal H}^{\prime }(i-1,i+1)
\right) = \exp (-\beta
                        ^{\prime }{\cal H}^{\prime }) \,\,\, ,
\end{eqnarray}
\narrowtext
where
\FL
\begin{eqnarray}
&&{\cal -}\beta {\cal H}(i,j) =
 -t \left( c_{i\sigma }^{\dagger }c_{j\sigma }+c_{j\sigma }^{\dagger
}c_{i\sigma}
                        \right) \\
&& -(U/2d)  \sum_i  \left( n_{i\uparrow}n_{i\downarrow }+n_{j\uparrow
}n_{j\downarrow }
                                \right )  +
\left( \mu /2d \right)
\sum_i \left( n_i+n_j \right). \nonumber
\end{eqnarray}
Thus, the approximation consists in
neglecting the commutation relations beyond segments of three consecutive
unrenormalized sites. This approximation is effected twice $(\simeq )$ in
Eq.(4), in opposing directions, hopefully with compensatory effect. The crux of
the calculation is extracted from the third step in Eq.(4),

\begin{equation}
{\bf Tr}_{w_2}
e^{-\beta {\cal H}(1,2) - \beta {\cal H}(2,3)}   =
e^{-\beta ^{\prime }{\cal H}^{\prime }(1,3)} \,\,\, .
\end{equation}
When written in terms of three-site (on the left) and two-site (on the right)
matrix elements, this equation amounts to
contracting a $64 \times 64$ matrix into a $16 \times 16$ matrix. This
operation is facilitated by
block diagonalization of the matrices, using the conservations of particles,
total spin magnitude, total spin $z$-component, and parity, so that the largest
blocks are $4 \times 4$ and $2 \times 2$ for the unrenormalized and
renormalized systems,
respectively. Thus, a renormalized Hamiltonian $-\beta ^{\prime }{\cal
H}^{\prime }$ is extracted.
The closed form of $-\beta ^{\prime }{\cal H}^{\prime }$ is more general
than Eq.(1),
namely
\widetext
\FL
\begin{eqnarray}
 -\beta {\cal H} = &&-\sum_{<ij>,\sigma}
        \left[
                t_0h_{i-\sigma}h_{j-\sigma } +
                t_1 \left( n_{i-\sigma }h_{j-\sigma }
                 + h_{i-\sigma }n_{j-\sigma} \right)
        + t_2n_{i-\sigma }n_{j-\sigma } \right]
                \left( c_{i\sigma }^{\dagger }c_{j\sigma
}+c_{j\sigma}^{\dagger }c_{i\sigma } \right) \nonumber \\
&&-t_x  \sum _{<ij>}
                \left( c_{i\uparrow }^{\dagger }c_{j\uparrow
}c_{i\downarrow }^{\dagger}
                c_{j\downarrow }+c_{j\uparrow }^{\dagger }c_{i\uparrow
}c_{j\downarrow}^{\dagger }
                c_{i\downarrow } \right) - U  \sum_i n_{i\uparrow
}n_{i\downarrow }+\mu
                \sum_i n_i \,\,   \\
&&+ \sum _{<ij>}
                \left[
                J\overrightarrow{s}_i\cdot
\overrightarrow{s}_j+V_2n_in_j+V_3 \left( n_{i\uparrow}n_{i\downarrow }
                n_j+n_in_{j\uparrow }n_{j\downarrow } \right)
                + V_4n_{i\uparrow}n_{i\downarrow }n_{j\uparrow
}n_{j\downarrow } \right], \nonumber
\end{eqnarray}
\narrowtext
\noindent
where the hole operator is
$h_{i\sigma }\equiv 1-n_{i\sigma }$ and the electron spin operator at site $i$
is
\begin{equation}
\overrightarrow{s}_i =  \sum_{\sigma,\sigma ^{\prime }} c_{i\sigma }^{\dagger}
\overrightarrow{s}_{\sigma \sigma ^{\prime }}c_{i\sigma ^{\prime }} \,\,\, ,
\end{equation}
where $\overrightarrow{s}_{\sigma \sigma ^{\prime }}$ is the vector of Pauli
spin matrices. The four hopping terms in the flow Hamiltonian [Eq.(7)]
correspond to one electron hopping with or without the opposite spin electron
present at the initial and final sites (two of these processes are related by
hermitivity and therefore have the same hopping strength $t_1$) and to two
electrons simultaneously hopping from one site to a neighboring site. These four
processes can be called vacancy hopping $(t_0)$, pair breaking $(t_1)$, pair
hopping $(t_2)$, and vacancy-pair interchange $(t_x)$. For
\begin{equation}
t_0 = t_1 = t_2\;,\;\;\;\;   t_x = J = V_2 = V_3 = V_4 = 0 \,\,\, ,
\end{equation}
the flow Hamiltonian [Eq.(7)] reduces to
the Hubbard Hamiltonian [Eq.(1)]. Thus, Eqs.(9) are the initial conditions of
our renormalization-group flows. However, in general, the hopping strengths
renormalize differently and the new interactions are generated under rescaling,
so that the renormalization-group flows are in the 10-dimensional,
$\overrightarrow{K} = (t_0,t_1,t_2,t_x,U,\mu ,J,V_2,V_3,V_4)$, Hamiltonian
space.

The transformation is implemented in $d>1$ by using the Migdal-Kadanoff
procedure, so that
$
\overrightarrow{K^{\prime }}=(b^{d-1}/f)\overrightarrow{R}(f\overrightarrow{K})
$
where $b=2$ is the length-rescaling factor, the function
$\overrightarrow{R}$ is the contraction process specified in the previous
paragraph, and $f$ is an arbitrary bond-moving factor, set to yield the correct
transition temperature of the Ising model ($f=1.2279$ and $1.4024 $ in $d=3$
and $2$). This renormalization-group transformation yields known information
about quantum systems, such as, in $d=1$, the absence finite-temperature phase
transitions; in $d=2$, a conventional phase transiton for the Ising model, a
Kosterlitz-Thouless transition for the XY model\cite{QXY1,QXY2}, no phase
transition for the Heisenberg model; in $d=3$, ferromagnetic and
antiferromagnetic phase transitions for the Heisenberg model, the
antiferromagnetic transition occurring at a 22\% higher temperature than the
ferromagnetic transition, a purely quantum mechanical effect\cite{SE}. The
10-dimensional renormalization-group flows also conserve the particle-hole
symmetry, given the map:
\FL
\begin{eqnarray}
\rule{.05in}{0in} \overline{t}_0 & = & t_2\;,\;\; \overline{t}_1 = t_1\;,
\;\; \overline{t}_2  =  t_0\;, \;\;
\overline{t}_x=t_x\;, \;\;
\overline{J} = J \,\,\, , \\
\rule{.05in}{0in} \overline{\mu  } & = & -\mu +U+2dV_3-2dV_4\;, \;\;\overline{U}
 =  U+4dV_3-2dV_4, \nonumber \\
\rule{.05in}{0in} \overline{V}_2 & = & V_2-2V_3+V_4\;, \;\;\overline{V}_3
 =  -V_3+V_4\;, \;\; \overline{V}_4=V_4.  \nonumber
\end{eqnarray}

The global analysis of the renormalization-group flows yields the phase diagram
of the system. We have thus obtained the global phase diagram of Hubbard model,
presented here in Figures 1--3, where first- and second-order phase
boundaries are
respectively shown by dottted and full curves. The particle-hole symmetry
[Eq.(10)] dictates that the Hubbard model [Eq.(1)] phase diagrams be symmetric
about $\mu /U=0.5$, which is seen in all of our results.

Figures 1 are for $U/t=20$. Figure 1(right panel) shows the full phase
diagram in temperature
versus chemical potential. Figures 1(left, middle panels) show the details
in temperature versus
electron density and chemical potential, respectively. It is seen that an
antiferromagnetic phase occurs around half-filling, purely due to electron
hopping, since the Hubbard Hamiltonian [Eq.(1)] does not contain an explicit
antiferromagnetic coupling. In fact, we traced the occurrence of this
antiferromagnetic phase to the non-zero value of the pair-breaking strength
$t_1$. The antiferromagnetic phase is unstable to at most 10\% hole or electron
doping from half filling. Between 30 to 40\% hole or electron doping, a $\tau $
phase occurs in which the vacancy hopping strength $t_0$ or the pair hopping
strength $t_2$ [see Eq.(7)], respectively, renormalizes to infinity under
repeated renormalization-group transformations. Thus, for hole doping, under
repeated renormalization-group transformations, $t_0\rightarrow \infty,J/t_0
= 2,V_2/t_0 = 3/2,\mu /t_0 =
6,t_{i\neq 0} = 0,U \rightarrow \infty , t_i/U = 0, V_i/U = 0$.
Symmetrically, for
electron doping, the overbarred variables of Eq.(10) have this behavior. In all
other regions of the phase diagram, all hopping strengths renormalize to zero
under repeated renormalization-group transformations. Near the $\tau $ phase, a
first-order phase transition (dotted curves) occurs, seen as a single curve in
Fig.1 (center) in terms of electron chemical potential and opening up into a
coexistence region in Fig.1 (left) in terms of electron density. The latter
shows
the distinctive feature of this first-order transition, namely that it involves
a very narrow phase separation, e.g., a discontinuity in electron density of
less than 2\%. This is similar to what is seen experimentally in lanthanide
compounds.\cite{lanth} At temperatures above the $\tau $ phase, a sequence of
antiferromagnetic and disordered phases is seen, at many temperature scales
[Figs.1,2 (center)]. We interpret this as the presence of an incommensurate spin
modulation phase, with a temperature- and (less strongly, by the alignment of
the sequencing) density-dependent periodicity. Our renormalization-group
transformation, with a commensurate rescaling factor and a built-in
approximation, acts as a spurious substrate potential which, at small
incommensuration, registers the incommensurate phase and, at large
incommensuration, disorders it. The incommensurate phase that we thus deduce is
indicated in Figs.1,2 (left). The features described above were also seen in the
simpler, less realistic, tJ model.\cite{FB,FBT}

As $U/t$ is decreased, the first-order phase boundary moves with respect to the
$\tau $ phase. It is seen that, for $U/t=4.44$ (Figs.2), it actually abuts the
boundary of the $\tau $ phase and, for $U/t=0.8$ (Figs.3), it is on the other
side of the $\tau $ phase.

We have thus calculated a finite-temperature phase diagram for the $d=3$ Hubbard
model that is rich in phase transition phenomena. We have also repeated the same
calculation for $d=2$. We find that no phase separation (in contrast to the tJ
model\cite{FB,FBT}) or other phase transition occurs at finite temperature
for the Hubbard model [Eq.(1)] in $d=2$.

This research was supported by the Italian Istituto Nazionale di Fisica Nucleare
(INFN), U.S. Department of Energy under Grant No. DE-FG02-92ER45473, and by the
Scientific and Technical Research Council of Turkey (T\"UBITAK). We gratefully
acknowledge the hospitality of the Feza G\"ursey Research Center for Basic
Sciences and of the Istanbul Technical University.

\begin{center}
{\bf Figure Captions}
\end{center}

\bigskip
\bigskip
\noindent
{\bf Figs.\ 1}: Calculated phase diagram of the $d=3$ Hubbard model for
$U/t=20$.
First- and second-order phase boundaries are shown with dotted and full
curves, respectively. Fig.1(right panel) shows the full phase diagram,
which is symmetric
about $\mu /U=0.5$. Antiferromagnetic [$a$], disordered [$D$], and  $\tau $
phases are seen. In the $\tau $ phases, the hopping strength $t_0$ or $t_2$
renormalizes to
infinity, for hole or electron doping respectively. Above the $\tau $
phase, a sequence of
antiferromagnetic and
disordered phases [Fig.1(center panel)] is interpreted as an incommensurate
spin modulation
phase [Fig.1(left panel)]. As seen in (left), the first-order phase
boundary has a very
narrow coexistence region.

\bigskip
\bigskip
\noindent
{\bf Figs.\ 2}:  Calculated phase diagram of the $d=3$ Hubbard model for
$U/t=4.44$.

\bigskip
\bigskip
\noindent
{\bf Figs.\ 3}. Calculated phase diagram of the $d=3$ Hubbard model for
$U/t=0.8$.

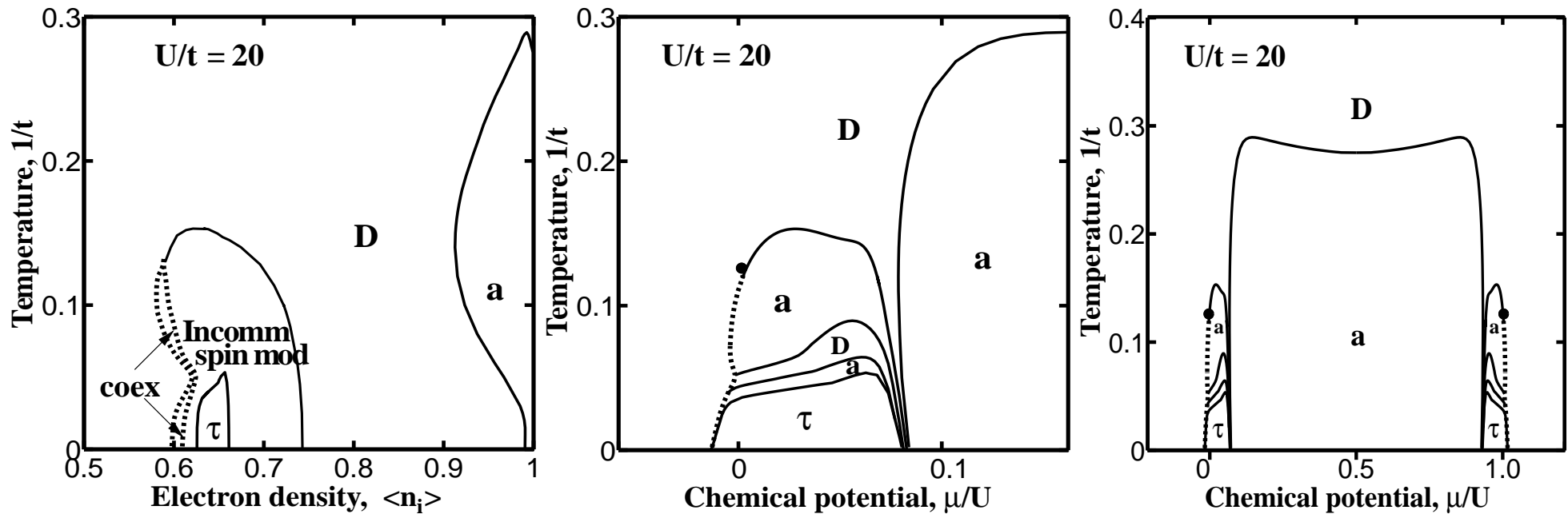
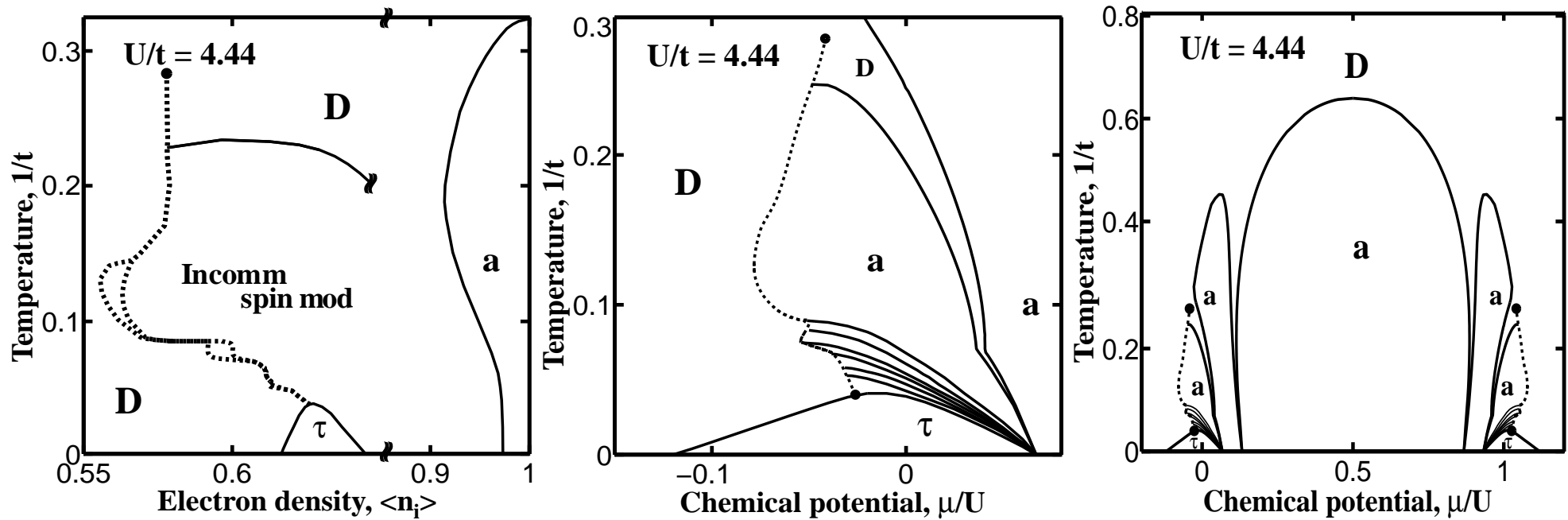
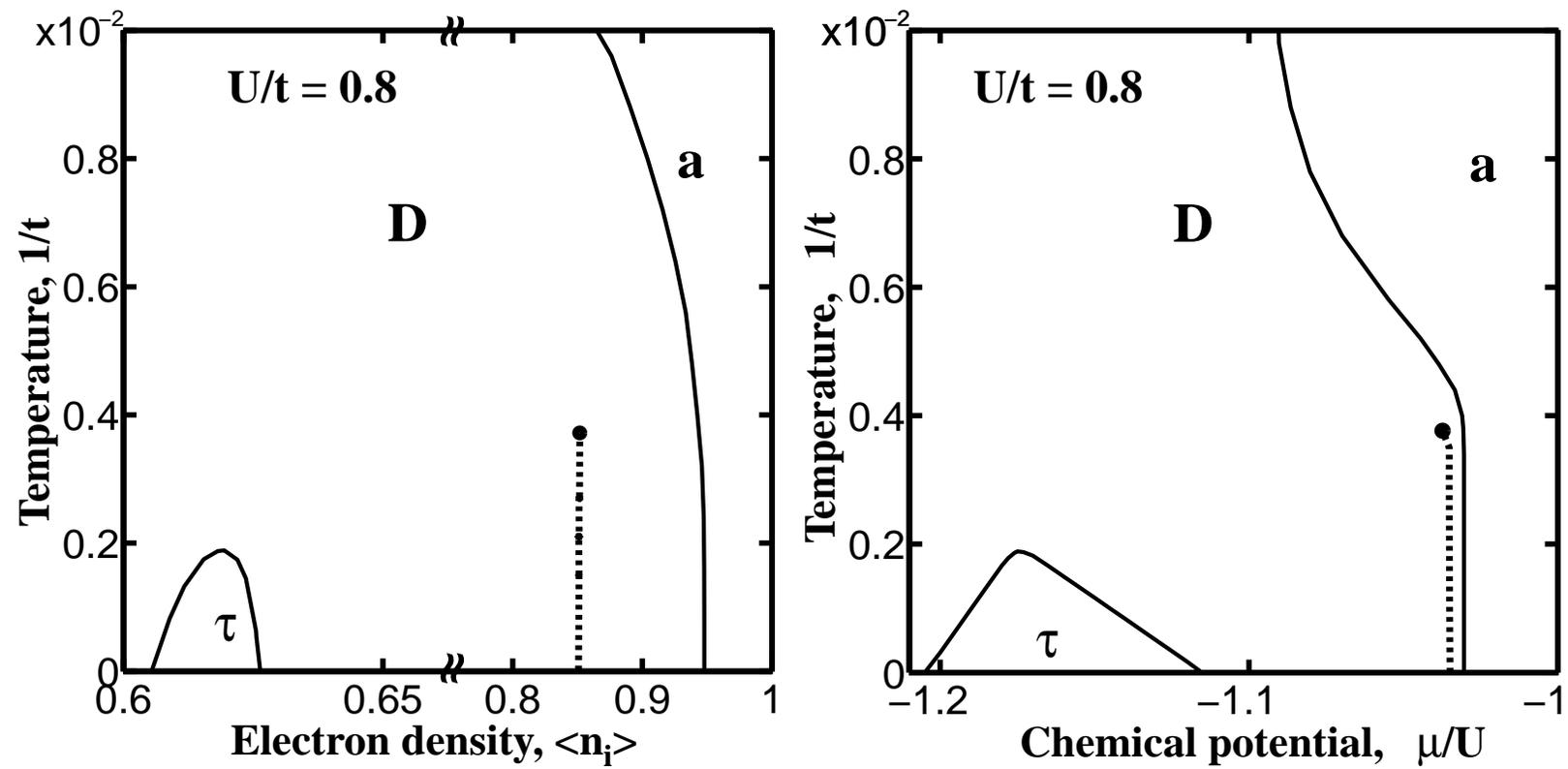
\end{document}